\documentclass[conference]{IEEEtran}

\IEEEoverridecommandlockouts

\usepackage{dblfloatfix}
\usepackage{placeins}

\usepackage{color}
\usepackage{xcolor}
\usepackage{amsthm, amsfonts}%, amssymb}
\usepackage{makecell}
\usepackage{textcomp}
\usepackage{epstopdf}
\usepackage{multirow}
\usepackage{wrapfig}
\usepackage{subfig}
\usepackage[ruled]{algorithm2e} % For algorithms 

\usepackage{bm}
\usepackage{enumitem}
\usepackage{balance}
\usepackage{graphicx}
\usepackage{amsmath}
\usepackage{cleveref}

\usepackage{booktabs}
\usepackage{tabularx}
\usepackage{multirow}
\usepackage{array}      % better p{}/X column control
\usepackage{makecell}   % optional: nicer line breaks in cells

\usepackage{amsthm}
\theoremstyle{plain}

\usepackage{algcompatible}
\usepackage{algorithmicx}

\usepackage{listings}
\usepackage{xcolor}

\newcommand{\ie}{\textit{i.e.}\xspace}

\newcommand{\Eq}[1]{Eq. (\ref{#1})}
\newcommand{\Fig}[1]{Fig.~\ref{#1}}

\newcommand{\Sec}[1]{Sec.~\ref{#1}}
\newcommand{\ignore}[1]{}

\begin{document}

\title{Skeletal-Anchored Dual Harmonics for Structured 3D Modeling}

\author{
\IEEEauthorblockN{
Zhentao Huang\IEEEauthorrefmark{1},
Changhao Li\IEEEauthorrefmark{2},
Ruizhen Hu\IEEEauthorrefmark{3},
Hui Huang\IEEEauthorrefmark{3},
and Minglun Gong\IEEEauthorrefmark{1}
}
\IEEEauthorblockA{\IEEEauthorrefmark{1}University of Guelph, Guelph, Canada\\
Email: \{zhentao, minglun\}@uoguelph.ca\\
Corresponding author: Minglun Gong}
\IEEEauthorblockA{\IEEEauthorrefmark{2}Math Magic, Beijing, China}
\IEEEauthorblockA{\IEEEauthorrefmark{3}Shenzhen University, Shenzhen, China}
}

\maketitle

\begin{abstract}
We present Skeletal-Anchored Dual Harmonics (SADH), a novel 3D shape representation that tightly couples local surface geometry with internal meso-skeletal organization. SADH represents a shape as a collection of compact surface patches rooted on internal anchors optimized directly inside the object volume. Each patch is parameterized using a dual-channel spherical harmonic (SH) formulation, where one channel models local radial geometry while the other defines adaptive patch support through a generalized viewing cone. Unlike isotropic primitives such as medial spheres or Gaussian kernels, SH patches directly encode anisotropic local surface geometry together with adaptive spatial support, enabling compact representation of detailed and directionally varying surface regions. Starting from unorganized point clouds, SADH jointly optimizes surface geometry, anchor locations, patch orientations, and structural connectivity through a staged optimization process that progressively forms a coherent meso-skeletal structure. A geodesic anchor graph further preserves structural relationships between neighboring patches. Experiments on complex 3D shapes demonstrate that SADH achieves accurate surface reconstruction together with compact and coherent skeletal organization across a wide range of geometries.
\end{abstract}


\begin{figure*}[t]
  \centering
  \includegraphics[width=\textwidth]{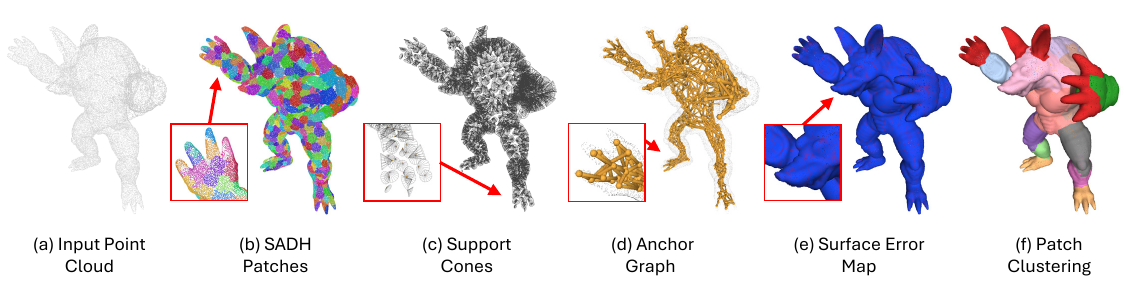}
  \caption{
Given an unorganized input point cloud (a), Skeletal-Anchored Dual Harmonics (SADH) decomposes the surface into compact dual-harmonic patches (b), each rooted at an internal anchor and defined over an adaptive support cone with learned height and half-angle (c). SADH also induces a geodesic anchor graph over the internal anchors (d), shown with the input points overlaid, which captures structural connectivity between neighboring patches. The reconstructed object surface is color-coded by reconstruction error (e), from low (blue) to high (red). The SADH patch parameters, together with the anchor graph, can be used to cluster patches into higher-level surface regions within a single shape or across multiple shapes (f).
  }
  \label{fig:teaser}
\end{figure*}

\section{Introduction}

3D shape representations lie at the core of geometry processing, reconstruction, and generative modeling. Recent advances in neural implicit fields, patch-based parameterizations, and primitive-based decompositions have substantially improved the fidelity and flexibility of digital shape modeling. However, many existing representations primarily focus on modeling external surface geometry, often without explicitly capturing the internal structural organization of shapes. As a result, reconstructed or generated surfaces may achieve strong geometric accuracy while lacking coherent structural abstractions that support downstream tasks such as deformation, animation, and topology-aware editing.

A long-standing direction in geometry processing addresses this problem through skeletal and medial representations. Classical approaches such as the L1-medial skeleton~\cite{L1Median} extract robust skeletal structures directly from point clouds, while Deep Points Consolidation (DPC)~\cite{DeepPoints} associates surface geometry with internal meso-skeletal structures composed of both curve-like and sheet-like components. More recent work, including Coverage Axis~\cite{Dou2022} and Coverage Axis++~\cite{Wang2024}, formulates skeletal abstraction through global coverage optimization. These approaches demonstrate the importance of internal structural reasoning for achieving compact and geometrically meaningful representations. However, they primarily focus on skeletal extraction or geometric consolidation rather than directly parameterizing continuous surface geometry.

In parallel, neural implicit and patch-based representations have emerged as powerful paradigms for shape modeling. Neural implicit fields such as DeepSDF~\cite{Park2019DeepSDF}, Occupancy Networks~\cite{Mescheder2019Occupancy}, and ConvONet~\cite{Peng2020Convolutional} model geometry through continuous occupancy or signed-distance functions, enabling high-fidelity reconstruction and generation. To improve locality and representation efficiency, subsequent methods explored patch-based and anchored parameterizations, including AtlasNet~\cite{Groueix2018AtlasNet}, ARO-Net~\cite{Wang2023_ARO}, and MASH~\cite{MASH}. These methods represent geometry using compact local surface patches rooted at spatial anchors. Primitive-based decomposition methods~\cite{Paschalidou2019Superquadrics,Deng2020CvxNet,Chen2020BSPNet,Paschalidou2021} further demonstrate the benefits of compact geometric abstractions for interpretable shape modeling. Nevertheless, many existing primitives remain isotropic or analytically constrained, making them less effective for representing thin structures and anisotropic local geometry.

We present Skeletal-Anchored Dual Harmonics (SADH), a novel 3D shape representation that tightly couples local surface geometry with internal meso-skeletal organization. SADH represents a shape as a collection of compact local surface patches rooted on internal anchors optimized directly inside the object volume. Each patch is parameterized using dual-channel spherical harmonics (SH), where one channel models local radial geometry while the other defines adaptive patch support through a generalized viewing cone. Unlike isotropic primitives such as medial spheres or Gaussian kernels, SH patches directly encode anisotropic local surface geometry together with adaptive spatial support, enabling compact representation of detailed and directionally varying surface regions.

Starting from unorganized point clouds, SADH jointly optimizes surface patches, anchor locations, and patch connectivity through a staged optimization to form a coherent meso-skeletal structure. Compared with existing anchored or implicit representations, SADH introduces three key advantages. First, it jointly models local geometry and internal structural organization within a unified representation. Second, the dual-channel SH parameterization compactly represents anisotropic surface regions with adaptive support. Third, the resulting geodesic anchor graph preserves structural relationships between neighboring patches, naturally supporting downstream applications such as deformation, animation, and unsupervised surface segmentation or co-segmentation; see \Fig{fig:teaser}.

In summary, our main contributions are:

\begin{itemize}[leftmargin=*,itemsep=1pt,topsep=2pt]
\item We introduce Skeletal-Anchored Dual Harmonics (SADH), a novel shape representation that organizes compact SH surface patches around an internal meso-skeletal structure.

\item We propose a staged optimization framework that jointly refines surface geometry, anchor locations, patch orientations, angular support, and geodesic structural connectivity directly from unorganized point clouds.

\item We introduce a dual-channel SH formulation that models both local radial geometry and adaptive patch support, enabling compact representation of anisotropic surface regions.

\item We demonstrate that SADH achieves accurate surface reconstruction together with coherent meso-skeletal organization across a wide range of complex geometries.
\end{itemize}

\section{Related Work}

Our work intersects with three closely related areas: primitive-based surface decomposition, implicit and patch-based shape modeling, and meso-skeletal structure extraction. We therefore organize the related work around these three themes, focusing on how each informs the design of SADH and where our representation differs.

\subsection{Primitive-Based Shape Decomposition}

Primitive-based representations approximate geometry using collections of compact geometric elements such as spheres, cuboids, superquadrics, or convex parts. Recent approaches including Superquadrics Revisited~\cite{Paschalidou2019Superquadrics}, CvxNet~\cite{Deng2020CvxNet}, BSP-Net~\cite{Chen2020BSPNet}, and Neural Parts~\cite{Paschalidou2021} employ learned geometric primitives for compact and interpretable shape decomposition.

These representations provide strong structural priors and compact parameterizations, but many rely on isotropic or analytically constrained primitives that can struggle to efficiently represent thin structures and anisotropic local geometry. Unlike isotropic primitives such as medial spheres or Gaussian kernels, SADH employs dual-channel spherical harmonic (SH) patches that directly encode anisotropic local surface geometry together with adaptive spatial support. This allows a single anchor to compactly represent geometrically detailed and directionally varying surface regions while preserving a compact structural representation.

\subsection{Implicit and Patch-Based Shape Modeling}

Neural implicit representations, including Signed Distance Functions (SDFs) and occupancy fields, have become a dominant paradigm for 3D surface modeling due to their flexibility and continuous formulation. Representative approaches such as DeepSDF~\cite{Park2019DeepSDF}, Occupancy Networks~\cite{Mescheder2019Occupancy}, and ConvONet~\cite{Peng2020Convolutional} encode geometry through globally defined neural fields that map spatial coordinates to occupancy or signed distance values. While these methods achieve high reconstruction fidelity, they typically require dense neural decoding for surface extraction and provide limited explicit structural organization.

To improve locality and representation efficiency, subsequent work explored patch-based and anchored surface representations. AtlasNet~\cite{Groueix2018AtlasNet} models shapes as collections of learnable surface patches, while ARO-Net~\cite{Wang2023_ARO} demonstrated the effectiveness of anchor-centered local geometry modeling for sparse-view reconstruction. MASH (Masked Anchored SpHerical Distances)~\cite{MASH} further introduced a compact parameterized representation based on spherical harmonic (SH) surface patches rooted at spatial anchors and constrained by generalized view cones. SADH inherits the compact local parameterization of anchored SH representations while extending them to an internally organized skeletal setting.

\subsection{Meso-Skeletal Structure Extraction}

The extraction of internal structural priors from 3D geometry has long been a central theme in shape analysis. A notable milestone is the L1-medial skeleton \cite{L1Median}, which leverages the statistical robustness of the L1-median to extract skeletal structures directly from raw point clouds without requiring normal information. Deep Points Consolidation (DPC) \cite{DeepPoints} further connected surface geometry with internal structure by associating surface points with internal points residing on a meso-skeleton composed of both curve-like and sheet-like structures \cite{L1Median, Tagliasacchi2012}. Through its sinking process, DPC encouraged points to settle into the object interior, providing a non-local structural prior for consolidating incomplete and noisy surfaces.

Beyond local consolidation-based formulations, recent work has explored global skeletal organization through coverage-based optimization. Coverage Axis \cite{Dou2022} and Coverage Axis++ \cite{Wang2024} formulate skeletal point selection as a set-cover problem, seeking a compact collection of internal primitives that jointly cover the entire surface. Medial Skeletal Diagram~\cite{guo2024medial} further generalizes medial-axis representations by encoding 3D shapes using compact skeletal diagrams, providing a structured internal abstraction for shape representation. Similar to these methods, SADH emphasizes global surface coverage to encourage compact and structurally consistent skeletal abstractions.

SADH extends these meso-skeletal perspectives from skeletal extraction and geometric consolidation to a fully parameterized surface representation. Instead of using unstructured point pairs or isolated skeletal primitives, SADH organizes compact SH surface patches around internal anchors that are jointly optimized with the surface representation itself. Surface geometry is modeled through point-to-patch assignments, while anchor connectivity is derived from the adjacency of their associated surface patches, forming a geodesic anchor graph that couples local geometry with internal structural organization.

\subsection{Point Cloud Segmentation and Part Discovery}

As an application of SADH to point cloud analysis, we demonstrate unsupervised shape segmentation and co-segmentation. Point cloud segmentation methods typically assign semantic or part labels directly to input points. Early neural approaches such as PointNet~\cite{Qi2017PointNet} and PointNet++~\cite{Qi2017PointNetPlusPlus} process unordered point sets directly, while graph-based and attention-based methods such as DGCNN~\cite{Wang2019DGCNN} and Point Transformer~\cite{Zhao2021PointTransformer} improve local feature aggregation for per-point prediction. These methods are usually supervised and operate directly at the point level. In contrast, our co-clustering experiment is unsupervised and operates on SADH patches: patch descriptors and the geodesic anchor graph are clustered first, and the resulting labels are transferred to surface points through the SADH assignment map. This highlights SADH as a structured intermediate representation for point-level organization, rather than a replacement for supervised segmentation networks.

\begin{figure*}[t]
    \centering
    \includegraphics[width=\linewidth]{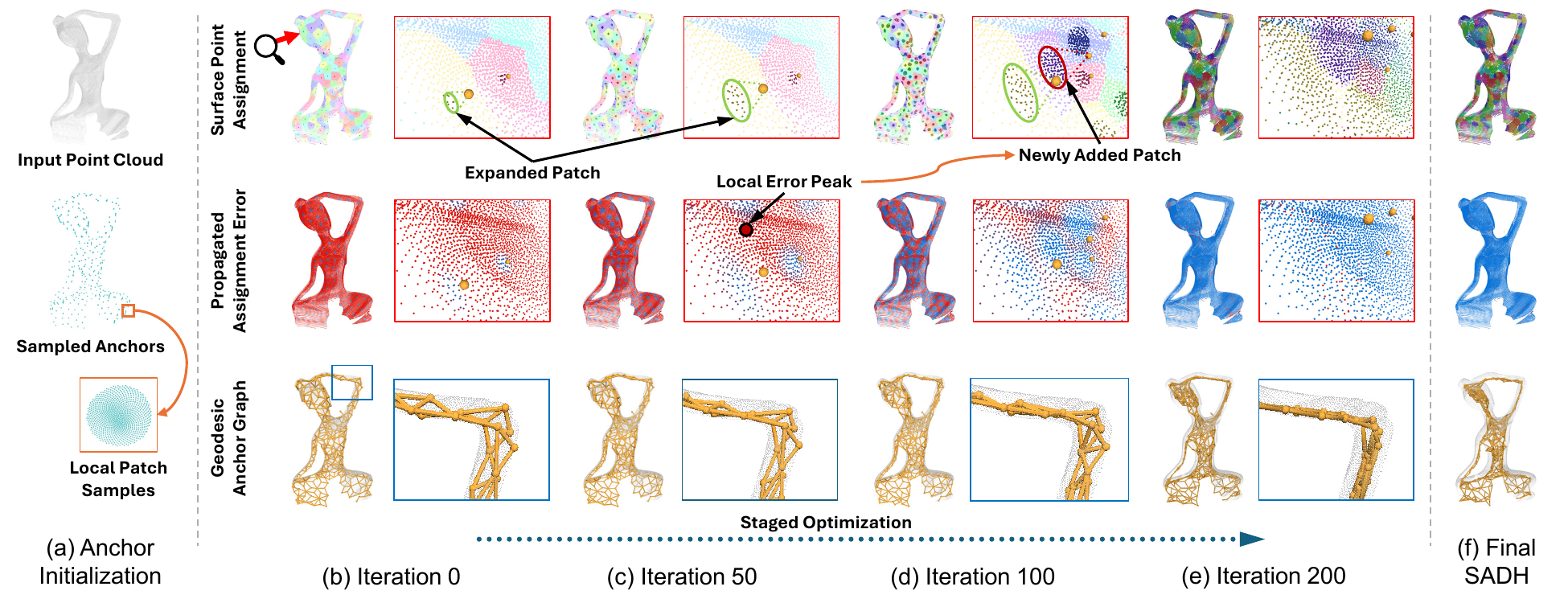}
    \caption{Algorithm overview. Starting from an unorganized and potentially incomplete input point cloud, SADH initializes sparse anchor points slightly beneath the surface and associates each anchor with a local SH surface patch sampled over canonical angular directions (a). During staged optimization, surface points are assigned to nearby patches through nearest-neighbor matching and graph-based propagation, producing propagated assignment errors and a geodesic anchor graph that captures structural connectivity between neighboring patches (b--f). Over iterations, anchors are pushed deeper into the object interior while the patches gradually expand, leading to reduced assignment errors (b to c). High local error regions may trigger adaptive anchor insertion (c to d). The process converges to a compact non-overlapping surface decomposition rooted on a coherent meso-skeletal structure.}
    \label{fig:overview}
\end{figure*}

\section{Skeletal-Anchored Dual Harmonics}

Given an unorganized input point cloud, SADH represents the shape as a collection of compact local surface patches organized around an internal meso-skeletal structure; see \Fig{fig:overview}. Each patch is anchored at an internal point and parameterized using a dual-channel SH representation, where the radial channel models local surface geometry and the alpha channel defines the patch support region through a generalized viewing cone.

Starting from sparse anchors initialized slightly beneath the observed surface, SADH performs a staged optimization process that jointly refines surface geometry, anchor locations, patch orientations, and angular support. Surface samples generated from the SH patches are matched to the input point cloud to establish patch assignments and propagated assignment errors through a graph-based propagation process. These assignments further induce a geodesic anchor graph that captures structural relationships between neighboring patches. During optimization, anchors gradually settle into the object interior to form a coherent meso-skeletal structure, while patches expand to improve surface coverage and reduce reconstruction error. Regions with persistently high local error trigger adaptive anchor insertion, progressively refining both geometric detail and structural organization. The optimization converges to a compact set of non-overlapping surface patches rooted on an internally organized skeletal structure.

\subsection{SADH Representation}

Skeletal-Anchored Dual Harmonics (SADH) represents a 3D shape as a set of continuous surface fields rooted on an internal meso-skeleton. Each anchor serves as a local generative center for a surface patch $P_i$, defined within a viewing cone centered at anchor location $\mathbf{a}_i\in\mathbb{R}^3$ and oriented by an outward-facing local frame $\mathbf{R}_i\in SO(3)$. The cone half-angle is controlled by an angular scale $s_i\in[0,1]$, where $s_i=1$ corresponds to a half-angle of $90^{\circ}$. Within this cone, patch $P_i$ is parameterized using dual-channel SH coefficients $\mathbf{Y}_i^{\langle r,\alpha\rangle}$, where the distance channel $r$ models the radial distance from the anchor to the surface and the alpha channel $\alpha$ defines the support boundary of the patch.

The entire patch is therefore represented as $P_i=\left(\mathbf{a}_i, \mathbf{R}_i, s_i, \mathbf{Y}_i^{\langle r,\alpha\rangle}\right)$, which has 42 parameters. It yields a valid sample at parameters $(\phi,\theta)$ if the support function $\mathbf{Y}_i^\alpha(\phi,\theta)$ exceeds a threshold. In that case, the sample location is computed as:
\begin{equation}
    \mathbf{y}_i(\phi,\theta) = \mathbf{a}_i + \mathbf{R}_i
    \left[
    \mathbf{Y}_i^r(\phi,\theta)\,
    \mathbf{u}(\phi, s_i\cdot \theta)
    \right].
    \label{eq:sample_location}
\end{equation}
where $\mathbf{u}(\cdot)$ maps spherical coordinates to a unit direction. The angular scale $s_i$ modulates the effective polar angle, thereby controlling the angular extent of the patch. As the anchor moves deeper into the interior of the object, $s_i$ decreases the cone half-angle accordingly, enabling more effective use of the SH representation. $\mathbf{Y}_i^\alpha$ further refines the active support of the surface patch based on local geometry and interactions with neighboring patches.

To impose structure on the surface patches, we further construct a geodesic anchor graph $\mathcal{G}=(\mathcal{A},\mathcal{E})$, where $\mathcal{A}=\{\mathbf{a}_i\}_{i=1}^M$ denotes the set of $M$ anchors and $\mathcal{E}$ encodes adjacency between anchors whose induced surface patches are spatially neighboring. This dual-harmonic formulation decomposes an unorganized point cloud into local, compact, oriented, and differentiable patches, while the geodesic anchor graph preserves the global structural organization of the object.

\subsection{SADH Optimization}
\label{sec:optimization}
Given an input shape represented by an oriented point set $\mathcal{S}=\{(\mathbf{x}_j,\mathbf{n}_j)\}_{j=1}^{N}$ sampled from a mesh or reconstructed surface, we optimize a set of SADH patches $\{P_i\}_{i=1}^M$ under three coupled objectives: (1) each patch $P_i$ accurately reconstructs its corresponding surface region; (2) the anchor points $\{\mathbf{a}_i\}_{i=1}^M$ form a coherent internal skeletal structure that captures the global topology of the object; and (3) the union of all patches $\bigcup_{i=1}^M P_i$ covers the entire surface.

Due to the large parameter space and the complexity of the optimization objectives, we adopt a staged optimization strategy rather than updating all parameters simultaneously. After patch initialization (\Sec{sec:initialization}), we iteratively perform surface point assignment (\Sec{sec:color_assignment}), dual-harmonic surface fitting (\Sec{sec:Harmonic_fitting}, which optimizes the SH parameters $\mathbf{Y}^{\langle r,\alpha\rangle}$), skeletal regularization (\Sec{sec:regularization}, which refines the anchors' locations $\mathbf{a}$), and coverage optimization (\Sec{sec:coverage}, which adjusts the anchors' orientations and angular scales $\{\mathbf{R}, s\}$).

\subsubsection{Anchor and Patch Initialization}
\label{sec:initialization}

We initialize anchors from the oriented input point set $\mathcal{S}$ by selecting surface points using Farthest Point Sampling (FPS) and moving each selected sample $(\mathbf{x}_i,\mathbf{n}_i)$ inward along its normal to initial anchor location $\mathbf{a}_i=\mathbf{x}_i-d_0\mathbf{n}_i$, where $d_0$ is an initial depth parameter. The local frame $\mathbf{R}_i$ is initialized such that its outward axis aligns with $\mathbf{n}_i$. We further set $s_i=0.2$ and $\mathbf{Y}_i^{\langle r,\alpha\rangle}$ to represent a spherical patch of radius $d_0$ with a half-angle of $36^\circ$. This yields the initial set of patches $\{P_i\}_{i=1}^M$.

To facilitate computation, we also construct a $k$-nearest neighbor ($k$NN) graph for the input points $\mathcal{S}$, which remains fixed throughout the optimization. Specifically, for each point $\mathbf{x}_j \in \mathcal{S}$, we denote its neighborhood as $\mathcal{N}_j \subset \mathcal{S}$, consisting of its $k$ nearest neighbors.

\subsubsection{Surface Point Assignment}
\label{sec:color_assignment}

Given the point set $\mathcal{S}=\{(\mathbf{x}_j,\mathbf{n}_j)\}_{j=1}^{N}$ and current set of patches $\{P_i\}_{i=1}^M$, we first assign each surface point $\mathbf{x}_j$ to a single patch $P_i$, so that each patch can be optimized based on its assigned points. That is, we define a labeling function $\ell: [N] \rightarrow [M]$, where $[K]:= \{1,\ldots,K\}$ denotes an index set.

The process starts by sampling each patch using a precomputed and fixed set of canonical angular directions, $\Omega=\left\{(\phi_q,\theta_q)\right\}_{q=1}^{Q}$, that uniformly sample the hemisphere.
Applying \Eq{eq:sample_location} yields a set of sample points $\mathcal{Y}_i=\{\mathbf{y}_{iq}\}_{q=1}^{Q}$ for patch $P_i$, where $\mathbf{y}_{iq}=\mathbf{y}_i(\phi_q,\theta_q)$. The complete SADH sample set over all $M$ patches is $\mathcal{Y}=\bigcup_{i=1}^{M}\mathcal{Y}_i$.

Next, we compute nearest-neighbor correspondences from $\mathcal{Y}$ to input point set $\mathcal{S}$. We denote by $c(i,q)=\arg \min_{j\in[N]} \|\mathbf{y}_{iq}-\mathbf{x}_{j}\|$ the index of the nearest input point to  sample $\mathbf{y}_{iq}$. These correspondences provide candidate assignments of surface points to patches. Specifically, we define
\begin{equation}
\mathcal{I}(j) = \{ i \in [M] \mid \exists q \in [Q],\ c(i,q) = j\},
\end{equation}
which denotes the set of patches that have samples mapping to the surface point $\mathbf{x}_j$. While our goal is to assign each surface point to a unique patch, ambiguities arise in practice: some input points may be adjacent to samples from multiple patches, \ie $|\mathcal{I}(j)| > 1$, while others may have no nearby samples, \ie $\mathcal{I}(j) = \emptyset$.

In the first scenario, where a surface point $\mathbf{x}_j$ is covered by more than one patch, we assign $\mathbf{x}_j$ to the patch most accurately represents it. The representation error of patch $P_i$ for point $\mathbf{x}_j$ is evaluated by first transforming $\mathbf{x}_j$ into the local frame $\mathbf{R}_i$ to obtain its spherical coordinates $(\phi_{ij},\theta_{ij})$. We then account for the angular scale $s_i$ by evaluating the SH surface along the normalized direction $(\phi_{ij},\theta_{ij}/s_i)$ and measuring the corresponding point-to-SH distance using:
\begin{equation}
    e(\mathbf{x}_j, P_i) =
    \|\mathbf{x}_j- \mathbf{y}_i(\phi_{ij},\frac{\theta_{ij}}{s_i})\|,
    \label{eq:error_represent}
\end{equation}
which is independent of the number of sampling directions $Q$. In contrast, previous approaches that rely on Chamfer distance to measure accuracy are sensitive to the sampling density \cite{MASH}.

The patch assignment for input point $\mathbf{x}_j$ and the corresponding assignment error is therefore computed as:
\begin{equation}
    \ell(j) = \arg\min_{i\in \mathcal{I}(j)}
    e(\mathbf{x}_j,P_i),
    \qquad
    e(j) = \min_{i\in \mathcal{I}(j)}
    e(\mathbf{x}_j,P_i).
    \label{eq:assign_direct}
\end{equation}

All input points assigned to $P_i$ in the above process form a \emph{direct point set}:
\begin{equation}
    \mathcal{J}_i^{\mathrm{dir}} = \{j \in [N] \mid \ell(j)=i \;\text{and}\; j \text{ is assigned via \Eq{eq:assign_direct}} \}.
\end{equation}
For visualization purposes, we assign each patch a random but unique color and color the set $\mathcal{J}_i^{\mathrm{dir}}$ accordingly; see \Fig{fig:overview}.

To handle surface points remain unassigned, we propagate assignments through the $k$NN graph. For an unassigned point $\mathbf{x}_k$, we search its neighboring dark points and assign it to the patch with the lowest propagated assignment error:
\begin{eqnarray}
    \ell(k) = \arg\min_{j \in \mathcal{N}_k}
    \left[
        e(j) +  \|\mathbf{x}_k-\mathbf{x}_j\|
    \right], \nonumber  \\ 
    e(k) = \min_{j \in \mathcal{N}_k}
    \left[
        e(j) +  \|\mathbf{x}_k-\mathbf{x}_j\|
    \right].
    \label{eq:assign_propagate}
\end{eqnarray}

The propagation proceeds iteratively until all surface points are assigned to a patch. When multiple patch labels propagate to the same surface point $\mathbf{x}_k$, the process selects the assignment that yields the lowest propagated assignment error, effectively minimizing a joint criterion that combines the source assignment error $e(j)$ and the geodesic distance between $\mathbf{x}_k$ and $\mathbf{x}_j$. We refer to these newly assigned points as propagated points and visualize them using colors of the same hue with increased lightness. That is:
\begin{equation}
\mathcal{J}_i^{\mathrm{pro}} = \{ j \in [N] \mid \ell(j)=i \} \setminus \mathcal{J}_i^{\mathrm{dir}}.
\end{equation}

The resulting point assignments also enable the construction of the geodesic anchor graph $\mathcal{G}=(\mathcal{A},\mathcal{E})$ through patch adjacency; see \Fig{fig:overview}. Specifically, if $j \in \mathcal{N}_k$ and $\ell(j) \neq \ell(k)$, we add an edge $(\ell(j), \ell(k))$ to $\mathcal{E}$. This naturally links anchors based on the geodesic proximity of their surface patches, rather than the Euclidean distance between the anchors.

\subsubsection{Dual-Harmonic Surface Fitting}
\label{sec:Harmonic_fitting}

Given the current surface point assignments and patch parameters, the next step is to optimize the dual-channel harmonic fields of each patch $P_i$ to better represent its assigned points. This is achieved by updating the support function $\mathbf{Y}_i^\alpha$ based on the surface region it supports and fitting the generated samples $\mathbf{Y}_i^r$ to the input point locations.

We start by optimizing $\mathbf{Y}_i^\alpha$ based on the surface point assignment results. Specifically, in regions where input points are assigned to patch $P_i$, $\mathbf{Y}_i^\alpha$ is encouraged to be 1, whereas in regions where input points are assigned to other patches, it is encouraged to be 0. Accordingly, we define the target value at the $q^{th}$ sample of patch $P_i$ as $\hat{\alpha}_{iq} = \mathbf{1}\left[\ell(c(i,q))=i\right]$ and optimize the alpha channel with a binary cross-entropy loss:
\begin{equation}
    \mathcal{L}_{\alpha}
    =
    -\frac{1}{|\mathcal{Y}|}
    \sum_{i,q}
    \left[
    \hat{\alpha}_{iq}\log \mathbf{y}^\alpha_{iq} +
    (1-\hat{\alpha}_{iq})
    \log(1-\mathbf{y}^\alpha_{iq})
    \right].
    \label{eq:loss_alpha}
\end{equation}

When computing the fitting loss, the $\alpha$ values are used as weights, since we aim to accurately represent only the surface points that are assigned to each patch. Specifically, we use a binary mask induced by the $\alpha$ channel:
\begin{equation}
    \mathcal{L}_{\mathrm{fit}}
    =
    \frac{
    \sum_{i,q}
    \mathbf{1}\!\left[
    \mathbf{y}^\alpha_{iq} \geq \tau_\alpha
    \right]
    \left\|
    \mathbf{y}_{iq}-\mathbf{x}_{c(i,q)}
    \right\|
    }{
    \sum_{i,q}
    \mathbf{1}\!\left[
    \mathbf{y}^\alpha_{iq} \geq \tau_\alpha
    \right]
    + \epsilon
    } ,
    \label{eq:loss_fitting}
\end{equation}

The dual-harmonic fitting objective $\mathcal{L}_{\mathrm{SH}}$ is therefore defined as a weighted sum of the above two losses. At this step, only the dual-channel SH coefficients $\mathbf{Y}^{\langle r,\alpha\rangle}$ are updated by minimizing $\mathcal{L}_{\mathrm{SH}}$
\begin{equation}
    \mathcal{L}_{\mathrm{SH}}
    =
    \lambda_{\mathrm{fit}}\mathcal{L}_{\mathrm{fit}}
    +
    \lambda_{\alpha}\mathcal{L}_{\alpha}.
    \label{eq:loss_harmonic}
\end{equation}

\begin{figure}[t]
    \centering
    \includegraphics[width=\linewidth]{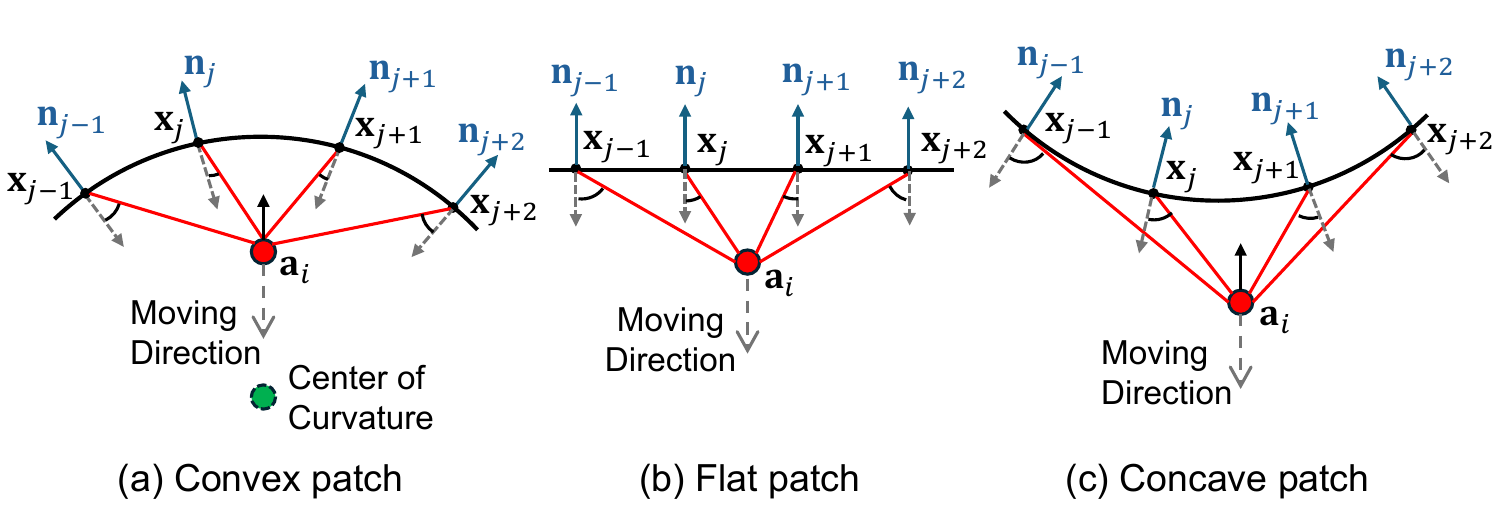}
    \caption{
    Illustration of the centering loss, which penalizes the angular deviation between the vector from a surface point $\mathbf{x}_j$ to its anchor $\mathbf{a}_i$ and the inward normal direction $-\mathbf{n}_j$. For a convex patch, this adaptively places the anchor near the center of curvature. For flat or concave surfaces, the constraint alone pushes the anchor toward infinity. 
    }
        \label{fig:centering_loss}
\end{figure}

\begin{figure}[]
    \centering
    \includegraphics[width=\linewidth]{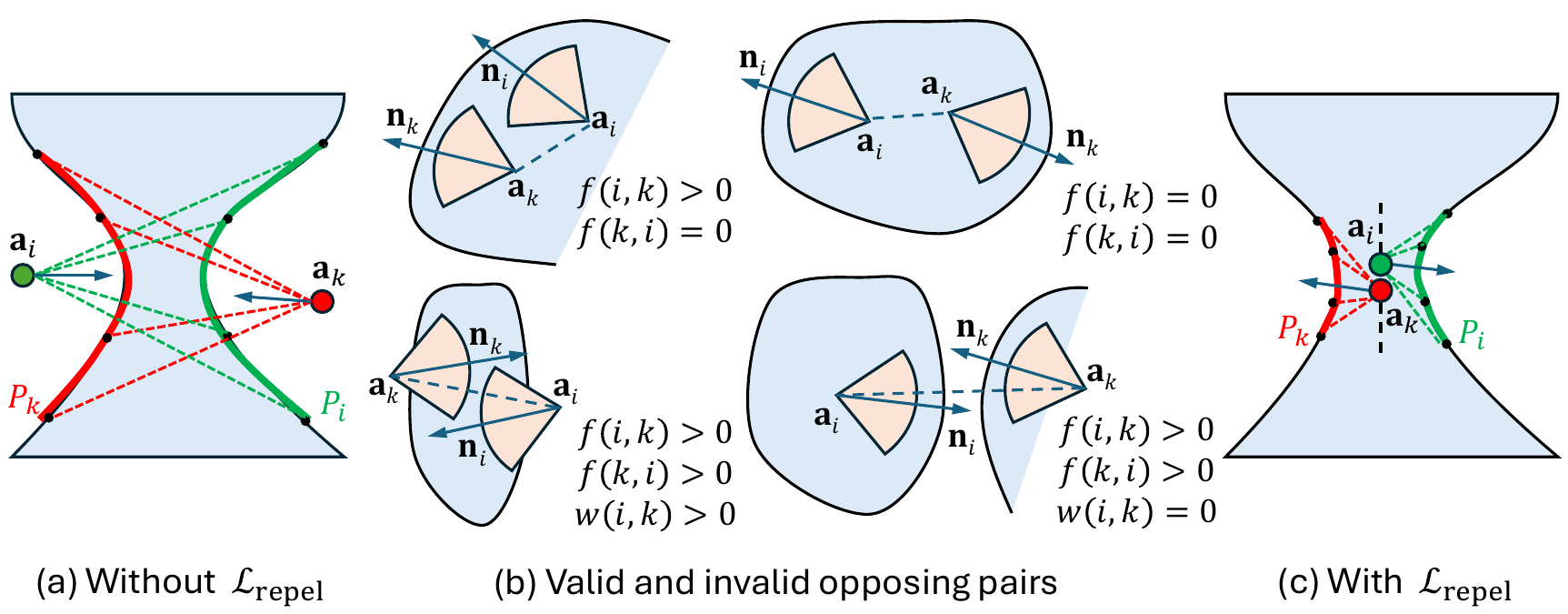}
    \caption{
    Illustration of the opposing-anchor repulsion loss. Left: in thin structures, anchors may pass through each other rather than forming a medial structure. Middle: the repulsion term is activated only for nearby anchors whose outward axes face each other, \ie, when all three terms in \Eq{eq:loss_repulsion} are non-zero. Right: the repulsion term encourages the anchors to settle on a common medial sheet or axis.
}
    \label{fig:repulsion_loss}
\end{figure}

\subsubsection{Skeletal Regularization}
\label{sec:regularization}

We now turn to the second objective discussed at the beginning of \Sec{sec:optimization}: optimizing the anchor locations to form a meso-skeleton that captures the object structure. This is a delicate process, for which we employ three loss terms.

The first term encourages each anchor to settle beneath the surface patch it generates. As illustrated in \Fig{fig:centering_loss}, this is achieved by encouraging the vector from each directly assigned surface point $\mathbf{x}_j$ to its anchor $\mathbf{a}_i$ to align with the inward normal direction, \ie, $\mathbf{n}_j^\top \frac{\mathbf{a}_i-\mathbf{x}_j} {\|\mathbf{a}_i-\mathbf{x}_j\|}=-1$. We therefore define:
\begin{equation}
    \mathcal{L}_{\mathrm{center}}
    =
    \frac{1}{M}
    \sum_{i=1}^{M} \left[
    \frac{1}{|\mathcal{J}^{\mathrm{dir}}_i|}
    \sum_{j\in\mathcal{J}^{\mathrm{dir}}_i}
    \left(
    1+
    \mathbf{n}_j^\top
    \frac{\mathbf{a}_i-\mathbf{x}_j}
    {\|\mathbf{a}_i-\mathbf{x}_j\|}
    \right) \right].
    \label{eq:loss_centering}
\end{equation}

The second term encourages each anchor $\mathbf{a}_i$ to lie on the meso-surface formed by its neighboring anchors in the geodesic anchor graph $\mathcal{G}=(\mathcal{A},\mathcal{E})$. Therefore, for each edge $(i,k)\in\mathcal{E}$, we penalize the signed distance between $\mathbf{a}_i$ and the tangent plane at its neighbor $\mathbf{a}_k$:
\begin{equation}
    \mathcal{L}_{\mathrm{meso}}
    =
    \frac{1}{|\mathcal{E}|}
    \sum_{(i,k)\in\mathcal{E}}
    \left|
    \mathbf{n}_k^\top
    (\mathbf{a}_i-\mathbf{a}_k)
    \right|.
    \label{eq:loss_projection}
\end{equation}

These two terms aim to position anchor points beneath their corresponding surface patches and along the meso-skeletal surface. However, for shapes with thin structures, anchors from opposing sides may pass through each other instead of converging to a common sheet or axis as illustrated in
\Fig{fig:repulsion_loss}. To address this issue, we construct a Euclidean neighborhood set $\mathcal{B}$ containing all anchor pairs $(i,k)$ whose spatial distance falls below a threshold, and define the following repulsion loss:
\begin{equation}
    \mathcal{L}_{\mathrm{repel}}
    =
    \frac{1}{\max(|\mathcal{B}|,1)}
    \sum_{(i,k)\in\mathcal{B}}
    f(i,k) f(k,i) w(i,k),
    \label{eq:loss_repulsion}
\end{equation}
where $f(i,k)=\left[(\mathbf{a}_k-\mathbf{a}_i)^\top \mathbf{n}_i \right]_+$ with $[\cdot]_+=\max(\cdot,0)$, measures whether anchor $i$ is facing anchor $k$, and $f(k,i)$ is defined symmetrically. The weighting term $w(i,k)=\left[\rho_{ik}-\|\mathbf{a}_i-\mathbf{a}_k\| \right]_+$ limits the penalty to nearby anchor pairs within an adaptive radius $\rho_{ik}$. The adaptive radius is defined as $\rho_{ik}=2\max(d_i,d_k)$, where $d_i=\mathbf{Y}_i^r(0,0)$ denotes the radial distance predicted along the central direction of patch $P_i$. This adapts the repulsion range to the local patch depth, preventing anchors associated with nearby but opposing surface patches from unnecessarily repelling each other; see \Fig{fig:repulsion_loss}(b).

Combining all three losses gives the objective for optimizing the anchor positions $\mathbf{a}$, encouraging anchors to settle onto a smooth meso-skeletal structure that preserves the patch connectivity induced by the target shape.
\begin{equation}
    \mathcal{L}_{\mathrm{skel}}
    =
    \lambda_{\mathrm{meso}}\mathcal{L}_{\mathrm{meso}}
    +
    \lambda_{\mathrm{center}}\mathcal{L}_{\mathrm{center}}
    +
    \lambda_{\mathrm{repel}}\mathcal{L}_{\mathrm{repel}}.
    \label{eq:loss_skeletal}
\end{equation}

\subsubsection{Coverage Optimization}
\label{sec:coverage}

Finally, we optimize the coverage of SADH surface patches, \ie, objective \#3. This is achieved by gradually adjusting the orientation and angular scale of individual patches $\{\mathbf{R},s\}$, removing patches that do not contribute sufficiently to the surface representation, and inserting new anchor points in poorly represented surface regions.

As shown in \Fig{fig:overview}, the initial set of anchor points is sparsely sampled and lies close to the surface, leaving large regions uncovered by the anchors' viewing cones. Input points in these uncovered regions are assigned to nearby anchors via propagation through \Eq{eq:assign_propagate}. While such assignments facilitate the construction of the geodesic anchor graph $\mathcal{G}=(\mathcal{A},\mathcal{E})$, we do not use them directly for optimizing surface patches. Instead, for each patch $P_i$, we construct a neighborhood-expanded set $\mathcal{J}_i^{\mathrm{nbr}}$ that includes all points in its direct point set $\mathcal{J}_i^{\mathrm{dir}}$ and their $k$NN neighbors. That is:
\[
\mathcal{J}_i^{\mathrm{nbr}}
=
\mathcal{J}_i^{\mathrm{dir}}
\cup
\bigcup_{k \in \mathcal{J}_i^{\mathrm{dir}}} \left(\mathcal{N}_k
\cap
\mathcal{J}_i^{\mathrm{pro}}\right)
\]

Next, we orient each anchor toward the center of its neighborhood-expanded set by minimizing the following orientation loss:
\begin{equation}
    \mathcal{L}_{\mathrm{orient}}
    =
    \frac{1}{M}
    \sum_{i=1}^{M} \left[
    \frac{1}{|\mathcal{J}^{\mathrm{nbr}}_i|}
    \sum_{j\in\mathcal{J}^{\mathrm{nbr}}_i}
    \frac{\theta_{ij}}{s_i} \right],
    \label{eq:loss_orientation}
\end{equation}
where $\theta_{ij}$ denotes the polar angle of $\mathbf{x}_j$ in the local frame $\mathbf{R}_i$, computed using the same transformation as in \Eq{eq:error_represent}. Minimizing $\mathcal{L}_{\mathrm{orient}}$ updates $\mathbf{R}_i$ and optimizes the orientations of the anchor points.

We further introduce a loss term that encourages the angular scale $s_i$ to increase so that the support region of patch $P_i$ expands to cover the neighborhood-expanded set. 
\begin{equation}
    \mathcal{L}_{\mathrm{coverage}}
    =
    \frac{1}{M}
    \sum_{i=1}^{M} \left[
    \frac{1}{|\mathcal{J}^{\mathrm{nbr}}_i|}
    \sum_{j\in\mathcal{J}^{\mathrm{nbr}}_i}
    \left[\frac{\theta_{ij}}{s_i}-\frac{\pi}{2}\right]_+
    \right].
    \label{eq:loss_coverage}
\end{equation}

Continuously enlarging the support region, however, reduces the representation capacity of the SH model. This may increase the representation error $e(\mathbf{x}_j, P_i)$ in regions with fine geometric detail, causing such points to be assigned to other patches. Therefore, we introduce a third loss term to reduce the number of surface points $\mathbf{x}_j$ that are covered by $P_i$ but assigned to other patches, \ie, $i \in \mathcal{I}(j) \wedge \ell(j) \neq i$.
\begin{equation}
    \mathcal{L}_{\mathrm{overlap}}
    =
    \frac{1}{N}
    \sum_{j=1}^{N}
    \sum_{i \in \mathcal{I}(j)}
    \mathbf{1}\!\left[\ell(j)\neq i\right]
    \left[
    \frac{\pi}{2}-\frac{\theta_{ij}}{s_i}
    \right]_+ .
    \label{eq:loss_overlap}
\end{equation}

Combining the coverage loss and overlap loss allows patches to gradually expand while avoiding excessive overlap, until the entire object surface is covered.
\begin{equation}
    \mathcal{L}_{\mathrm{scale}}
    =
    \mathcal{L}_{\mathrm{coverage}}
    +
    \lambda_{\mathrm{overlap}}
    \mathcal{L}_{\mathrm{overlap}}.
    \label{eq:loss_scale}
\end{equation}

Complex 3D shapes often require dense anchor placement in regions with fine geometric detail and sparse anchors in smooth areas. Such adaptive sampling cannot be achieved through the initial FPS-based anchor generation. To enable adaptive anchor placement, we periodically perform anchor deletion and insertion during optimization. Specifically, after 80 iterations, we identify anchor points whose colors are assigned to fewer than 5 surface points. Since all anchors compete during surface point assignment, this indicates that nearby patches provide better representations of the local surface. Therefore, these anchors can be safely removed.

Next, we evaluate the representation error $e(\mathbf{x}_j)$ over all surface points $\mathbf{x}_j$ and select the top 20 points with the highest error using non-local suppression. A new anchor is then initialized at each selected surface point $(\mathbf{x}_k,\mathbf{n}_k)$ by placing it at $\mathbf{a}_k=\mathbf{x}_k-\mathbf{n}_k^\top(\mathbf{x}_k-\mathbf{a}_i)$, where $\mathbf{a}_i$ is the location of the anchor that currently colors $\mathbf{x}_k$. The new anchor frame $\mathbf{R}_k$ is aligned with the normal $\mathbf{n}_k$, and its dual-channel harmonic $\mathbf{Y}_k^{\langle r,\alpha\rangle}$ is initialized to represent a spherical patch of radius $\mathbf{n}_k^\top(\mathbf{x}_k-\mathbf{a}_i)$ with a half-angle of $36^\circ$.

With angular scales optimized at each iteration and anchor points adaptively deleted and inserted every 40 iterations, we then return to the surface point assignment step to reevaluate the correspondence between surface points and anchors, followed by high-frequency SH parameter optimization and lower-frequency anchor refinement in subsequent rounds.

\section{Experiments}
We evaluate SADH in three settings: surface reconstruction with skeletal structure recovery, robustness to incomplete point clouds, and unsupervised cross-shape patch co-clustering. Together, these experiments assess SADH as both a compact surface representation and a structured intermediate representation for shape analysis.
\subsection{Surface Reconstruction and Skeletal Structure}

We first evaluate SADH on surface reconstruction from oriented point clouds using the Famous dataset from Points2Surf~\cite{erler2020points2surf}, which is widely used in reconstruction studies including PGR~\cite{lin2022surface} and WNNC~\cite{lin2024fast}.
The dataset contains 22 well-known geometry-processing models spanning organic characters, animals, and articulated shapes. Its rich structural components, thin regions, and branching geometries make it suitable for evaluating whether SADH can induce meaningful skeletal-anchor configurations while preserving accurate surface geometry.

For each shape, we uniformly sample an oriented point cloud $\mathcal{S}=\{(\mathbf{x}_j,\mathbf{n}_j)\}_{j=1}^{N}$ with $N=40{,}000$ points from the reference mesh and use it as input. After optimization, SADH produces a set of dual-harmonic surface patches rooted on skeletal anchors. For quantitative evaluation, we densely sample points from the reconstructed SADH surface and compare them against points sampled from the reference mesh. To obtain an explicit mesh representation, we decode the optimized SADH patches into dense oriented surface samples, which are then passed to a Poisson-style surface reconstruction backend to extract the final mesh. Reconstruction quality is evaluated using Chamfer distance, F-score, and normal consistency.

We compare SADH with WNNC~\cite{lin2024fast}, a strong point-cloud reconstruction baseline, and MASH~\cite{MASH}, the closest anchored parametric representation baseline. All methods use the same input point clouds with 40k oriented samples. For MASH and SADH, we decode the optimized representations into dense oriented samples and use the same meshing pipeline.

% \begin{table}[t]
%     \centering
%     \caption{
%     Quantitative surface reconstruction results on the Famous dataset with 40k input points.
%     Chamfer distance is multiplied by $10^3$ for readability.
%     }
%     \label{tab:famous_reconstruction}
%     \small
%     \setlength{\tabcolsep}{4pt}
%     \begin{tabular}{lcccccc}
%         \toprule
%         Method 
%         & L1-CD $\downarrow$ 
%         & L2-CD $\downarrow$ 
%         & F-score $\uparrow$ 
%         & $D_H$ $\downarrow$ 
%         & $S_{\cos}$ $\uparrow$ 
%         & NIC $\downarrow$ \\
%         \midrule
%         % PGR~\cite{lin2022surface} 
%         %     & -- & -- & -- & -- & -- & -- \\
%         WNNC~\cite{lin2024fast} 
%             & 4.357 & 3.508 & 0.9690 & 0.01853 & 0.9575 & 10.553 \\
%         MASH~\cite{MASH} 
%             & 7.643 & 6.813 & 0.8165 & 0.03591 & 0.8849 & 20.162 \\
%         SADH (Ours) 
%             & 6.095 & 5.395 & 0.9003 & 0.02876 & 0.9146 & 15.881 \\
%         \bottomrule
%     \end{tabular}
% \end{table}
\begin{table}[t]
    \centering
    \caption{Quantitative surface reconstruction results on the Famous dataset with 40k input points. Chamfer distance is multiplied by $10^3$ for readability.}
    \label{tab:famous_reconstruction}
    \scriptsize
    \setlength{\tabcolsep}{2.5pt}
    \renewcommand{\arraystretch}{1.05}
    \resizebox{\columnwidth}{!}{
    \begin{tabular}{lcccccc}
        \toprule
        Method & L1-CD $\downarrow$ & L2-CD $\downarrow$ & F-score $\uparrow$ & DH $\downarrow$ & $S_{\cos}$ $\uparrow$ & NIC $\downarrow$ \\
        \midrule
        WNNC~\cite{lin2024fast} & 4.357 & 3.508 & 0.9690 & 0.01853 & 0.9575 & 10.553 \\
        MASH~\cite{MASH}        & 7.643 & 6.813 & 0.8165 & 0.03591 & 0.8849 & 20.162 \\
        SADH (Ours)             & 6.095 & 5.395 & 0.9003 & 0.02876 & 0.9146 & 15.881 \\
        \bottomrule
    \end{tabular}
    }
\end{table}

Table~\ref{tab:famous_reconstruction} reports quantitative results on the Famous dataset. WNNC achieves the best surface metrics, as expected for a dedicated reconstruction method. SADH substantially improves over MASH across all measures, reducing L1-CD from 7.643 to 6.095, improving F-score from 0.8165 to 0.9003, and improving normal consistency from 0.8849 to 0.9146. This shows that optimizing SH patches around internal anchors improves coverage and local fidelity over an external anchor-based formulation.

\begin{figure*}[p]
    \centering
    \includegraphics[width=\linewidth]{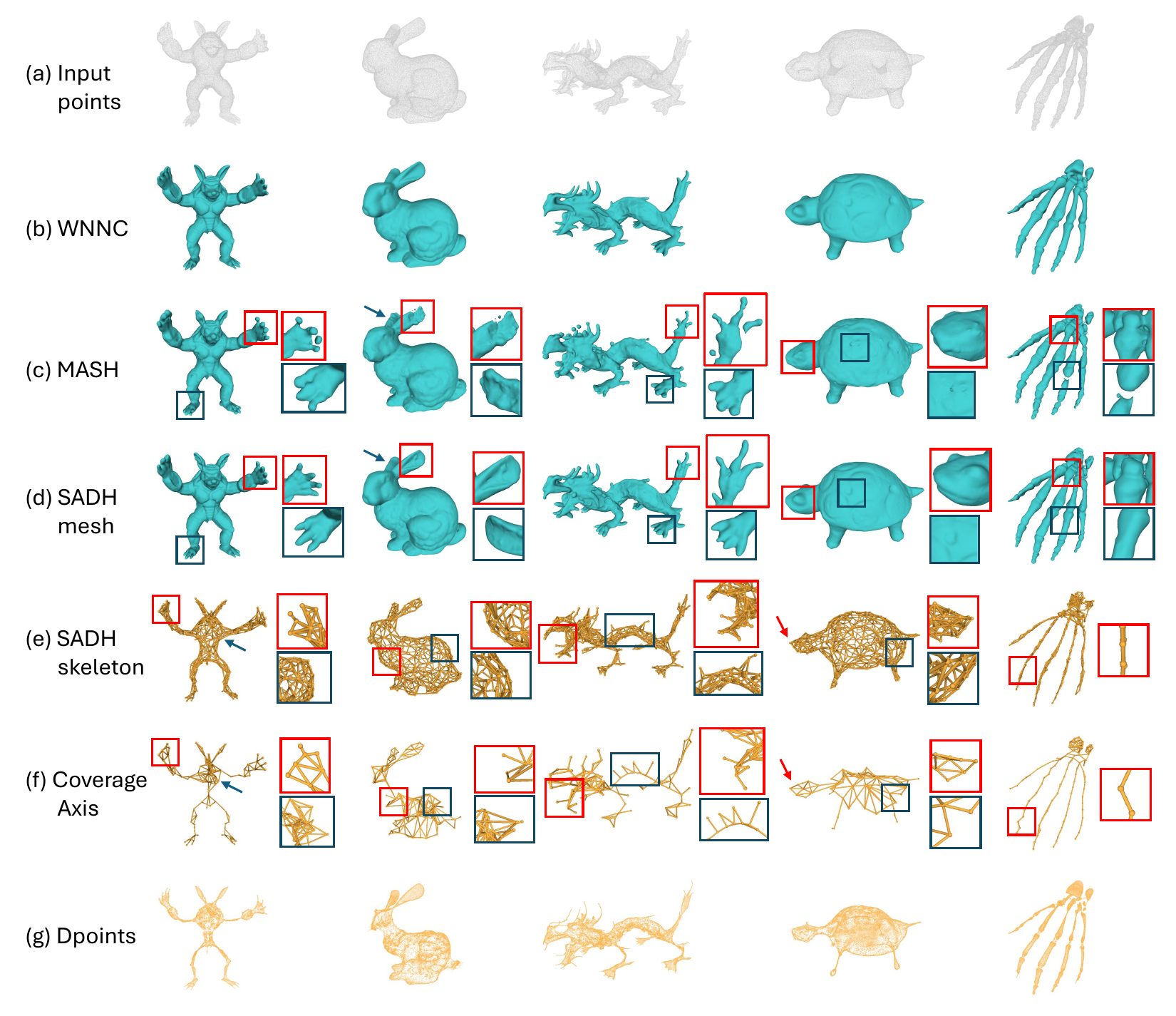}
    \caption{
Qualitative comparison on the Famous dataset. Given the same input point clouds (a), WNNC (b) reconstructs smooth surfaces but no internal structure, while MASH (c) may lose details or introduce artifacts around thin and articulated regions. SADH reconstructs comparable surfaces (d) and simultaneously produces a geodesic anchor graph (e) aligned with major structural components. Compared with Coverage Axis (f) and Dpoints (g), the SADH skeleton is directly tied to optimized surface patches, yielding a coherent meso-skeletal organization.
    }
    \label{fig:reconstruction_comparison}
\end{figure*}

Figure~\ref{fig:reconstruction_comparison} provides qualitative comparisons of reconstructed surfaces and internal structures. WNNC produces accurate surfaces without structural abstraction, while MASH may lose details or introduce artifacts around thin and articulated regions. SADH jointly reconstructs the surface and a geodesic anchor graph. Compared with Coverage Axis~\cite{Dou2022} and Deep Points~\cite{DeepPoints}, its structure is optimized together with the surface patches and remains directly associated with the reconstructed geometry.

\subsection{Robustness to Incomplete Point Clouds}

\begin{figure*}
    \centering
    \includegraphics[width=.8\linewidth]{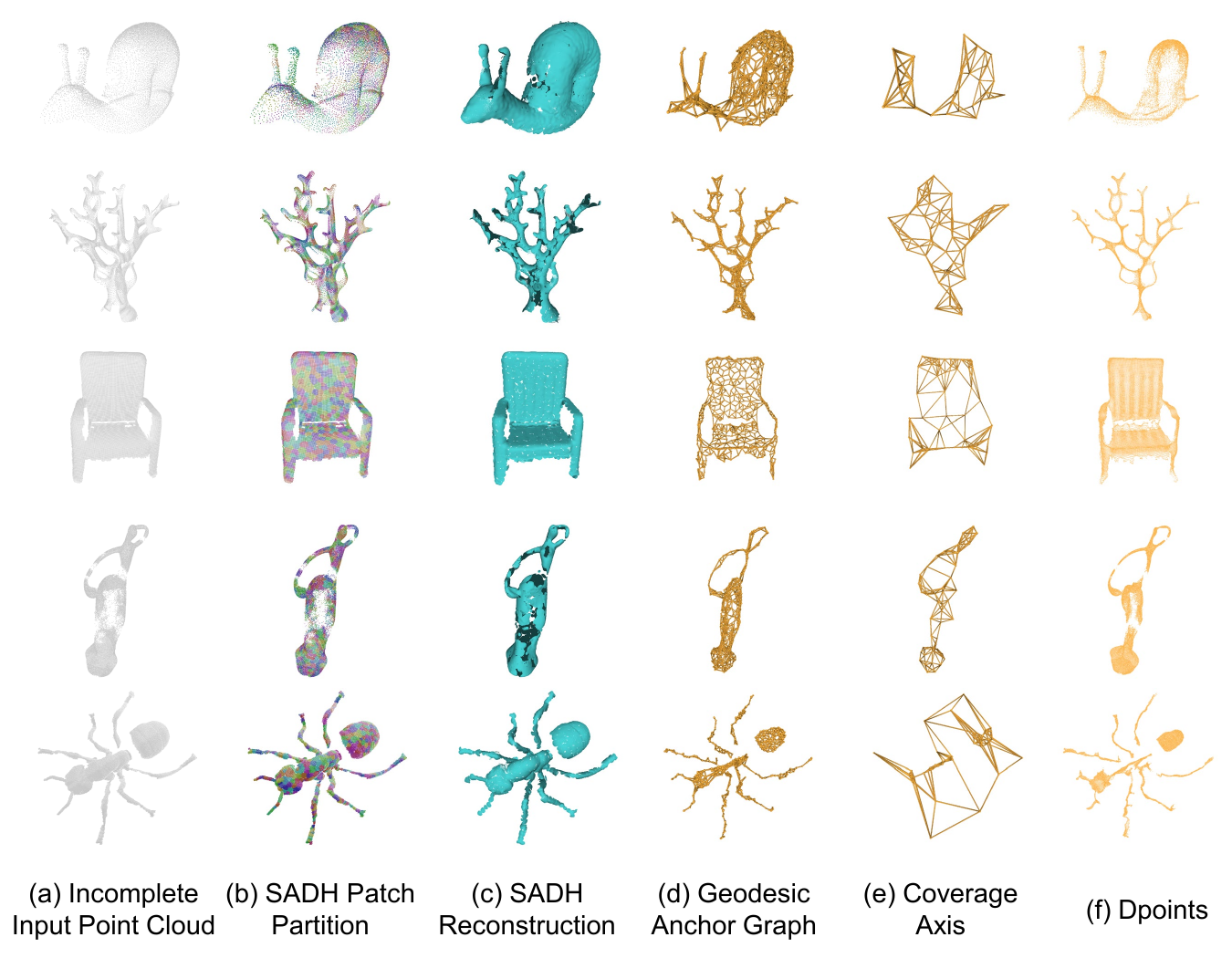}
    \caption{
Reconstruction and structural organization from incomplete point clouds. Given partial observations (a), SADH produces adaptive surface patches (b), reconstructs local surfaces (c), and forms a geodesic anchor graph aligned with major structures (d). Compared with Coverage Axis (e) and Dpoints (f), SADH yields a surface-aware meso-skeletal organization directly tied to the reconstructed patches.}
    
    \label{fig:incomplete_point}
\end{figure*}

We further evaluate SADH on incomplete input point clouds in \Fig{fig:incomplete_point}. SADH decomposes partial observations into coherent surface patches and reconstructs plausible local surfaces from the optimized dual-harmonic fields. More importantly, the induced geodesic anchor graph remains aligned with major structural components even when parts of the surface are missing. In contrast, Coverage Axis can introduce spurious branches and long unstable connections in missing or sparsely sampled regions. Dpoints provides point-to-point links between surface samples and skeletal points, producing smoother internal structures but often requiring dense skeletal samples. SADH instead establishes point-to-patch associations, allowing fewer anchors to represent larger coherent surface regions, which can simplify downstream deformation. By coupling skeletal organization with patch optimization, SADH recovers structures that remain directly associated with the reconstructed surface rather than only the sparse input samples.

\subsection{Unsupervised Cross-Shape Patch Co-Clustering}

Finally, we demonstrate SADH as an intermediate representation for unsupervised cross-shape part discovery. Since each SADH patch provides both a compact geometric descriptor and a node in the geodesic anchor graph, patches from multiple shapes can be clustered jointly without semantic supervision. For each patch, we extract a descriptor containing its anchor location, angular scale, dual-channel SH coefficients, and relative distances and orientations to neighboring anchors. We then construct a global patch graph over all shapes: intra-shape edges are given by the geodesic anchor graph, while inter-shape edges connect patches with similar descriptors across shapes. Spectral clustering~\cite{Ng2001Spectral} is applied to this graph, and the resulting patch labels are transferred back to surface points through the SADH assignment function.

\begin{figure*}
    \centering
    \includegraphics[width=.8\linewidth]{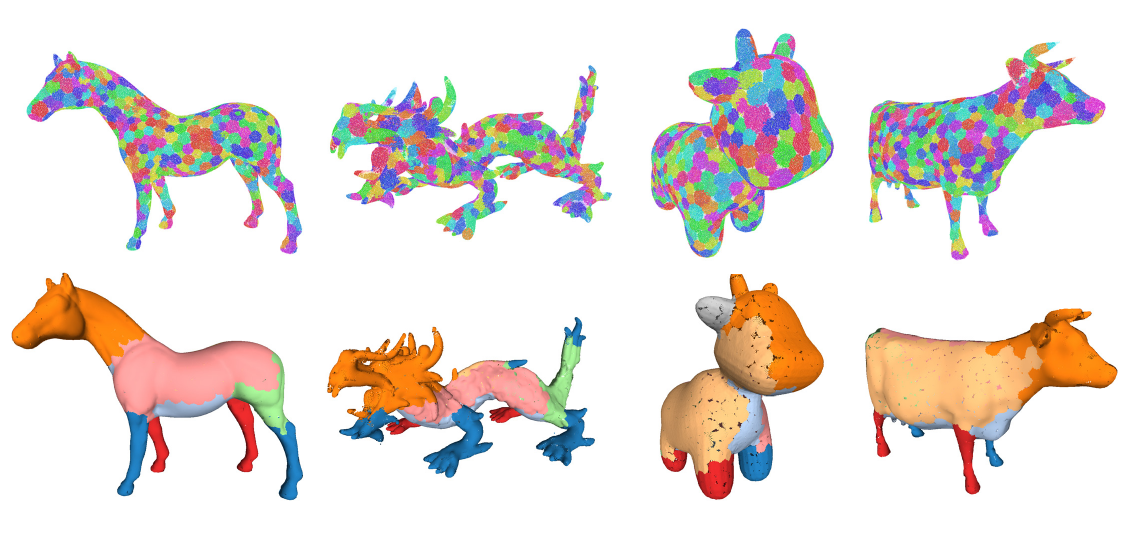}
    \caption{
Unsupervised cross-shape patch co-clustering using SADH. Patch descriptors and geodesic anchor graph connectivity are used to build a global graph across shapes, followed by spectral clustering. Without semantic supervision, the method discovers recurring structures across different poses and proportions, with similar colors corresponding to related regions such as body, limbs, head, and tail.    
    }
    \label{fig:coclustering}
\end{figure*}

Figure~\ref{fig:coclustering} shows co-clustering results on animal shapes. Despite variations in pose, proportion, and sampling, the discovered clusters remain largely consistent across models, grouping recurring structures such as limbs, body, head, neck, and tail-like regions with similar colors. Since no semantic labels are used, the results are not intended as fully semantic segmentation. Instead, they show that SADH provides a meaningful intermediate representation for grouping geometrically and structurally similar surface regions across shapes. This highlights the advantage of coupling local patch descriptors with a geodesic anchor graph, allowing clustering to consider both local surface appearance and structural context.

\section{Conclusion}

We presented Skeletal-Anchored Dual Harmonics (SADH), a compact 3D shape representation that couples local surface modeling with internal meso-skeletal organization. SADH represents an input shape as a collection of dual-channel SH patches rooted on internal anchors, where the radial channel models local surface geometry and the alpha channel defines adaptive patch support. Through staged optimization, SADH jointly refines patch geometry, anchor positions, orientations, angular scales, and geodesic anchor connectivity directly from unorganized point clouds.

Experiments show that SADH improves over the closest anchored SH baseline in surface reconstruction while producing coherent internal structures aligned with object geometry. It remains robust under incomplete observations, where the learned geodesic anchor graph stays associated with reconstructed surface patches rather than sparse input samples. We further show that SADH supports downstream tasks such as unsupervised cross-shape patch co-clustering, suggesting its potential as an intermediate representation for structured shape analysis.

While SADH provides a compact and structurally organized representation, it still depends on optimization quality, patch initialization, and the chosen SH degree. Future work includes extending SADH to category-level learning, integrating learned patch descriptors, and exploring deformation, animation, and generative modeling applications built on the induced meso-skeletal structure.

% Put remaining qualitative figures before references.
% This prevents Figs. 5--7 from appearing after the bibliography.
\FloatBarrier
\clearpage

\clearpage

\bibliographystyle{IEEEtran}
\bibliography{SADH}

\end{document}